\documentclass{ws-procs9x6}
\usepackage{graphicx}
\usepackage{rotating}
\begin{document}

\title{RELATIVISTIC SIGNATURES AT THE GALACTIC CENTRE}

\author{E. M. Howard}

\address{Faculty of Science and Engineering,\\Department of Physics and Astronomy,\\Macquarie University,\\
Sydney, NSW, Australia\\
$^*$E-mail: katie.howard@mq.edu.au\\
www.mq.edu.au}

\begin{abstract}
The current studies of the inner few parsecs at the Galactic Centre
provide enough indications of a supermassive black hole, object
associated with an unusual, variable (radio, near-infrared, and X-ray) source Sgr A*. The highly compact
nature of the distributed mass as well as the intense gravitational field suggest
the indisputable presence of dark matter.

There is also evidence that the main emissions from
Sgr A* originate from the accretion disc from within ten gravitational radii
from the dynamical centre. In order to understand the physics behind the observed
time-dependent physical phenomena, we study light curves and spectra of
emissions originated at the surface of the accretion disk, close to the event
horizon, near the marginally stable orbit of a Kerr (rotating) black hole. 

Our main goal is the investigation and deep analysis of the physical processes 
responsible for the variable observed emissions from the compact radio source Sgr A*.

\end{abstract}


\bodymatter

\section{Introduction} \label{aba:sec1}

Due to its relative proximity, SgrA* provides favorable circumstances 
to a better understanding of the processes responsible for the observed 
time-dependent phenomena.

Apparently, the emissions from Sgr A* are originated in  
radiative processes in keplerian motion, with a peak occuring within several
Schwarzschild radii ($r_S\equiv 2GM/c^2$) of the centre. 

Latest theoretical models are trying to 
address the question of what the spectral line profiles or continuum
may tell us about the black hole properties, and
analyze the constraining parameters and characteristics of the accretion disc
close to the event horizon. We take into account emissions from within 
the last stable orbit of a rotating (Kerr) black hole 
or from the region near above
the marginally stable orbit

Using fully relativistic ray-tracing methods, we analyze the constraining
parameters and principal characteristics of the black hole and study the main
imprints of both special and general relativistic effects on the time-resolved
emisssion phenomena.

We consider all relativistic effects (energy shift, aberration, light bending, lensing and relative time delay) near a Kerr black hole.
General and special relativistic effects play a crucial role in the
time-dependent behaviour, particularly for a maximally rotating Kerr black
hole. 

By integrating the photon geodesic paths between a position
inside a spot located within the accretion disk and an observer positioned at infinity,
we obtain time dependent spectra for an orbiting or infalling spot co-moving 
with the accretion disk, in a deep gravitational potential.

A detailed time-resolved analysis makes possible the study of the
evolution of the emitting region, close to the event horizon as well as the diagnosis of the
light curves of the variable emission region, for various spectral emissivity
profiles, different viewing directions of the distant observer and different locations
of the spot relatively to the local observer, the event horizon and the
center of the disk.

\section{Basic assumptions and objectives} 
We consider a complete system comprising a black hole, an accretion disc and a
co-rotating spot within the cold accretion disk. 

The gravitational field is described in terms of
Kerr metric, for a rotating black hole and particularly for a non-rotating black hole. Therefore, both static
Schwarzschild and rotating Kerr black hole cases are considered. 

The
co-rotating Keplerian accretion disc is geometrically thin and optically
thick, therefore we take into account only photons coming from the
equatorial plane directly to the observer. The spot is orbiting within
the disk near the rotating black hole.

We also assume that the
matter in the accretion disc is cold and neutral.

A fully
relativistic ray tracing code is used to
analyze how the considered effects affect and modify the Sgr A* disk emissions 
originated in a co-moving (local) frame with the accretion disk, up to a distant observer,
located at infinity. The observer is placed in the azimuthal direction $\varphi = \pi/2$. 

Relativistic effects alter the shape of the spectra, 
more significantly at large inclination angles and play an important role
in the time-dependent emission processes, especially in the case of a 
maximally rotating Kerr black hole.

\section{Summary}

We try to address various questions concerning the validity of the simple spot model and
the role of the relativistic effects in the observed phenomena.
We calculate time-dependent spectra of both cases of an orbiting or a free-falling spot.

\begin{figure}
\begin{center}
\includegraphics[width=12cm]{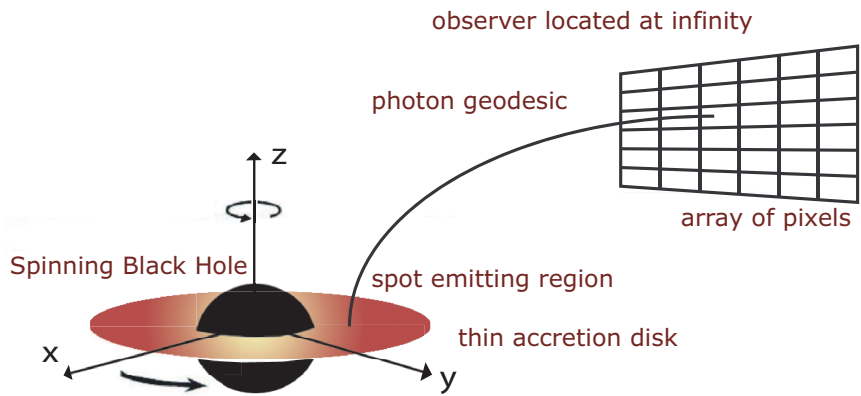}
\end{center}
\caption{Ray tracing in Kerr geometry}
\end{figure}

The code integrates
the local emission in polar coordinates on the disc and, as a consequence, 
we may handle emission originated in a non-axisymmetric area of integration.

In the axisymmetric case, the local emission is integrated in one
dimension, the radial coordinate of the disc.

\begin{figure}
\begin{center}
\includegraphics[width=11cm]{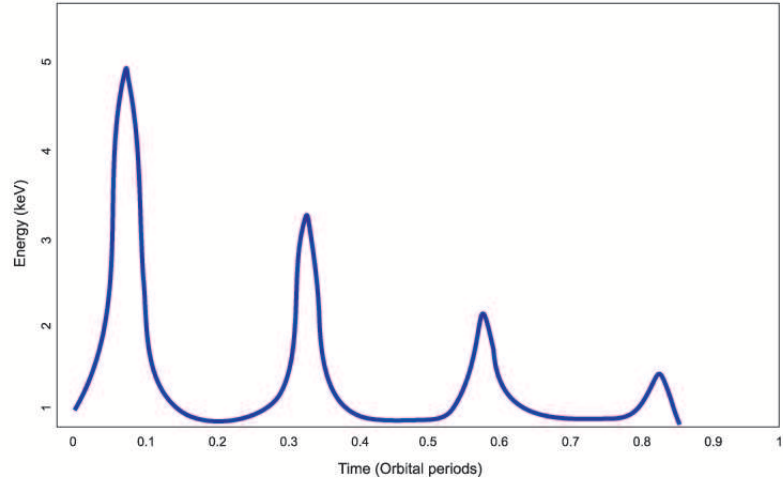}
\end{center}
\caption{Spectrum of an in-falling spot near to the horizon of an extreme Kerr black hole.
with spin parameter a=0.998. Energy is on the ordinate, time on the abscissa.}
\end{figure}

We are able to:  

1) gather information about the black-hole spin

2) study free-falling photons from the marginally stable orbit down
towards the black hole horizon. This provides us with information about 
the plunge region, the area between the Event Horizon and the last stable orbit

3) work with the emissivity as dependent on both of the polar coordinates within the equatorial plane

4) deal with time variable spectra, therefore handle non-stationary cases, making possible the study of the time evolution of the emitting region. 

5) study light curves for different inclination angles of the observer relative to the disk axis ($\theta_{\rm{}o}$) and analyze the dependence of the variability on the inclination.

6) choose a non-axisymmetric emission area, therefore control the size and shape of the spot 

7) consider non-axisymmetric geometry of accretion flows

8) analyze the general and special relativity effects that influence
photons paths along null geodesics towards an observer located at infinity

9) as the photon paths are integrated in Kerr ingoing
coordinates, this allows us to study the Kerr geometry and test the metric

Emission starts within a localized spot on the accretion disc. We consider
two cases, whether the spot moves with a Keplerian velocity along a stable circular orbit or, if close 
to the marginally stable orbit, it plunges into the black hole.

The intrinsic intensity at each point depends on the energy shift of the photons and it is time dependent.

Photons emerging from the spot area
would be affected by all relativistic imprints. 
The KYSPOT code by Dovciak et al.
takes all special and general relativistic effects into account
by using pre-calculated sets of transfer functions that map various properties of the emission region in
the accretion disk onto the sky plane (Cunningham 1975, 1976).

The transfer function, obtained by integration of the geodesic equation, correlates the flux in the local frame
comoving with the disk, to the flux as seen by an observer
located at infinity. 

Spectral line profiles are obviously affected by relativistic smearing but due to the power-law character
of the relativistic imprints, the main shape of the primary continuum profile remains intact.

Only the central gravity potential influences the trajectories of the photons the towards the observer. 

The functions are calculated for different values of
observer inclination angle and black hole horizons. 

The binary extensions contain information for different
radii and different values of $g$-factor, defined as 
the ratio between the photon energy observed at infinity
and the local photon energy as emitted from the disc.

\begin{figure}
\begin{center}
\includegraphics[width=11cm]{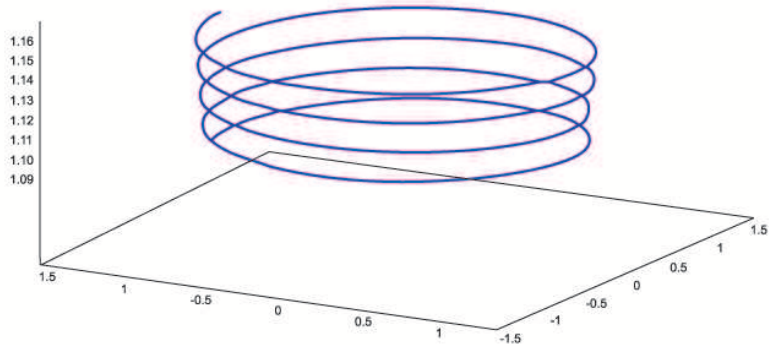}
\end{center}
\caption{Plunging trajectory of a photon into the event horizon of an extreme Black Hole}
\end{figure}

We obtain time-dependent spectra for various
viewing directions of a distant observer based on different emissivity profiles and various angular momenta of the black hole.

The relativistic corrections on the local emission 
are parameterized by the black hole spin 
and the observer's inclination angle.

The intrinsic emissivity is specified in the frame co-moving with the disc
medium and it is defined as a function of $r$, $\varphi$
and $t$ in the equatorial plane. When the geodesic integration is ended, after the transfer of photons to the distant observer is performed, Boyer-Lindquist coordinates will replace the initial Kerr ingoing coordinate system.

The observed spectra depend on the position of the spot with respect
to the disk normal.
We obtain light curves of the variable emission region, for different position and angles (azimuthal and polar) of the spot relatively to the distant observer, the event horizon and the center of the disk. We also consider   different spin parameter values, various viewing angles and different sizes of the emitting spot.

\section{References and literature}

\end{document}